\documentclass[vecphys]{svmult}

\usepackage{makeidx}         
\usepackage{graphicx}        
\usepackage{multicol}        
\usepackage[bottom]{footmisc}

\makeindex             

\begin{document}

\title*{Semileptonic and nonleptonic decays of $B_c$}
\author{
Mikhail A. Ivanov\inst{1} \and
J\"{u}rgen G. K\"{o}rner\inst{2} \and
Pietro Santorelli\inst{3}}
\institute{
Bogoliubov Laboratory of Theoretical Physics,
Joint Institute for Nuclear Research, 141980 Dubna, Russia
\and
Institut f\"{u}r Physik, Johannes Gutenberg-Universit\"{a}t,
D-55099 Mainz, Germany
\and
Dipartimento di Scienze Fisiche, Universit{\`a} di Napoli "Federico II",
Istituto Nazionale di Fisica Nucleare, Sezione di Napoli, Italy
\texttt{Pietro.Santorelli@na.infn.it}
\footnote{Preprint DSF-2006/27 (Napoli)}
}
%
%
\maketitle
\begin{abstract}
Using our relativistic constituent quark model we present results on the exclusive
nonleptonic and semileptonic decays of the $B_c$-meson.
The nonleptonic decays are studied in the framework of the
factorization approximation. We calculate the
branching ratios for a large set of exclusive nonleptonic and
semileptonic decays of the $B_c$ meson and compare our results with the
results of other models.
\end{abstract}

\section{Introduction}
\label{sec:intro}

The $B_c$-meson is the lowest bound state of
two heavy quarks (charm and bottom) with open
flavor. The $B_c$-meson therefore decays weakly via
(i) b-quark decay, (ii) c-quark decay, and (iii) the annihilation channel.
Starting from the pioneering paper \cite{{Lusignoli:1990ky}},
the modern state of art in the spectroscopy, production and decays of the
$B_c$-meson can be found in the review
\cite{Brambilla:2004wf}.

The first observation of the $B_c$ meson was reported by the CDF Collaboration
at Fermilab \cite{CDF} in the semileptonic decay mode  $B_c\to J/\psi +l +\nu$
with the $J/\psi$ decaying into muon pairs. Values for the mass and the
lifetime of the $B_c$ meson were given as $M(B_c)=6.40\pm 0.39\pm 0.13$ GeV
and  $\tau(B_c)=0.46^{+0.18}_{-0.16}({\rm stat})\pm 0.03({\rm syst})$ ps.
Recently, CDF reported new value for the mass of $B_c$ meson,
$6.2857 \pm 0.0053({\rm stat.}) \pm 0.0012 ({\rm syst.})$ GeV
with errors significantly smaller than in the first
measurement. Also D0 has observed the $B_c$ in the semileptonic
mode $B_c\!\to\! J/\psi+\mu +X$ and reported preliminary evidence
that $M(B_c)\! =\! 5.95^{+0.14}_{-0.13} \pm 0.34$ GeV
and $\tau(B_c)\!=\! 0.45^{+0.12}_{-0.10}\!\pm\! 0.12$ ps \cite{Corcoran:2005ti}.

In the following we report on the results of an analysis of
almost all accessible low-lying exclusive nonleptonic two-body
and semileptonic three-body modes of the $B_c$-decays \cite{last}
within our relativistic constituent quark model
\cite{Ivanov:2000aj,Faessler:2002ut,Ivanov:2002un,Ivanov:2005fd}.
In \cite{last} we updated the free parameters of the model by using the latest
experimental data on the $B_c$-mass \cite{Acosta:2005us}
and the weak decay constant $f_D$ \cite{Artuso:2005ym}. We give a set of numerical values
for the leptonic, semileptonic and nonleptonic partial decay widths
of the $B_c$-meson and compare them with the results of other approaches.

\section{Results and discussions}
\label{s:mod}

The constituent relativistic quark model we employ to study $B_c$ decays
was developed  in \cite{Ivanov:2000aj,Faessler:2002ut,Ivanov:2002un,Ivanov:2005fd}
and successfully applied to a very large class of weak decays (see for example
\cite{IvanovSantorelli}). For technical details regarding the model we refer the interested reader to ref \cite{last}.
Here we present our results on the semileptonic decays into charmonia,
into ($\overline B_s^0,\, \overline B_s^{0\ast}, \overline D^0,\,
\overline D^{0\ast},\, \overline B^0,\, \overline B^{0\ast}$)(cf. table \ref{tab:Bc-semlep1}) and
on the two body nonleptonic $B_c$ decays (cf. tables \ref{tab:Bc-nonlep1} and \ref{tab:Bc-nonlep2}).
Tables \ref{tab:Bc-nonlep1} and \ref{tab:Bc-nonlep2}
contain numerical results corresponding to
processes with branching ratios larger than
$0.1\%$. For the complete list see the tables in \cite{last}.

\begin{table}
\centering
\caption{\label{tab:Bc-semlep1}
Branching ratios (in $\%$) of exclusive semileptonic $B_c$ decays. $\psi$ indicates $\psi(3836)$. $\tau(B_c)=0.45$ ps.}
\def\arraystretch{1}
\begin{tabular}{ll}
\begin{tabular}{|l|l|l|l|l|l|l|}
\hline
 Mode & \cite{last}  & \cite{KKL,exBc} & \cite{Chang:1992pt}  & \cite{Faust} & \cite{AbdEl-Hady:1999xh} & \cite{Nobes:2000pm} \\
\hline
$B_c^- \to \eta_c e \nu$     & 0.81 & 0.75 & 0.97 & 0.40 & 0.76 & 0.51 \\
$B_c^- \to \eta_c \tau \nu$  & 0.22 & 0.23 & -    & -& -& -\\
\hline
$B_c^- \to J/\psi e \nu $    & 2.07 & 1.9  & 2.35 & 1.21 & 2.01 & 1.44 \\
$B_c^- \to J/\psi \tau \nu $ & 0.49 & 0.48 & -     &- &- & -\\
\hline
 $B_c^- \to  \overline D^0 e \nu $       & 0.0035    & 0.004 & 0.006   & 0.001 & 0.003 & 0.0014\\
 $B_c^- \to  \overline D^0 \tau \nu $  & 0.0021    & 0.002 & -       & -& -& -\\
\hline
 $B_c^- \to  \overline D^{\ast 0} e \nu  $ & 0.0038    & 0.018 & 0.018    & 0.008 & 0.013 & 0.0023 \\
 $B_c^- \to  \overline D^{\ast 0} \tau \nu$& 0.0022    & 0.008 & -        & - & - & -\\
\hline\hline
 $B_c^- \to  \overline B^0_s e \nu  $  & 1.10    & 4.03  & 1.82    & 0.82 & 0.98 & 0.92 \\
 $B_c^- \to \overline B_s^{\ast 0} e \nu$  & 2.37    & 5.06  & 3.01   & 1.71 & 3.45 & 1.41\\
\hline
 $B_c^- \to \overline B^0 e \nu  $    & 0.071    & 0.34  & 0.16   & 0.04 & 0.078 & 0.048\\
 $B_c^- \to \overline B^{\ast 0} e \nu  $  & 0.063    & 0.58  & 0.23    & 0.12 & 0.24 & 0.051\\
\hline
\end{tabular}\hspace{-.15truecm}
&  
\def\arraystretch{1.32}
\begin{tabular}{|l|l|l|}
\hline
 Mode & \cite{last}  & \cite{Chang:2001pm} \\
\hline
$B_c^-\to \chi_{c0} e\nu$     & 0.17   & 0.12  \\
$B_c^-\to\chi_{c0} \tau\nu$   & 0.013  & 0.017  \\
\hline
$B_c^-\to \chi_{c1} e\nu$     & 0.092   & 0.15 \\
$B_c^-\to \chi_{c1} \tau\nu$  & 0.0089  & 0.024\\
\hline
$B_c^-\to h_c e\nu $          & 0.27   & 0.17 \\
$B_c^-\to h_c \tau \nu$       & 0.017  & 0.024\\
\hline
$B_c^-\to \chi_{c2}e \nu $    & 0.17   & 0.19 \\
$B_c^-\to \chi_{c2}\tau \nu$  & 0.0082 & 0.029 \\
\hline
$B_c^-\to \psi e\nu $   & 0.0066 & - \\
$B_c^-\to \psi \tau\nu$ & $9.9\!\!\times\!\! 10^{\!-\!5}$& -\\
\hline
\end{tabular}
\end{tabular}
\end{table}

\begin{table}[t]
\centering
\caption{\label{tab:Bc-nonlep1}
         Branching ratios (in $\%$)
         of exclusive nonleptonic $B_c$ decays
         with the choice of Wilson coefficient:
         $a_1^c =1.20$ and $a_2^c=-0.317$ for c-decay,
         and $a_1^b =1.14$ and $a_2^b=-0.20$ for b-decay. Modes with branching ratios
         smaller than $0.1 \%$ can be found in table V of ref \cite{last}.}
\def\arraystretch{1}
\begin{tabular}{|l|l|l|l|l|l|l|}
\hline
 Mode & This work  & \cite{KKL,exBc} & \cite{Chang:1992pt} & \cite{Faust} & \cite{AbdEl-Hady:1999xh} \\
\hline
 $B_c^- \to \eta_c \pi^-$  & 0.19   & 0.20  & 0.18  & 0.083 & 0.14  \\
 $B_c^- \to \eta_c \rho^-$ & 0.45   & 0.42  & 0.49   & 0.20 & 0.33  \\
\hline
 $B_c^- \to J/\psi \pi^-$  & 0.17   & 0.13  & 0.18  & 0.060 & 0.11 \\
 $B_c^- \to J/\psi \rho^-$ & 0.49   & 0.40  & 0.53   & 0.16 & 0.31  \\
\hline
 $B_c^- \to \eta_c D_s^{-}$  & 0.44  & 0.28 & 0.054  & -& 0.26  \\
 $B_c^- \to \eta_c D_s^{\ast\,-}$ & 0.37  & 0.27 & 0.044 &  -& 0.24  \\
 $B_c^-  \to J/\psi D_s^-$    & 0.34  & 0.17 & 0.041 &  -& 0.15  \\
 $B_c^-  \to J/\psi D_s^{\ast\,-}$ & 0.97  & 0.67 & -     &  -& 0.55  \\
\hline\hline
 $B_c^- \to \overline B_s^0 \pi^-$     & 3.9  & 16.4 & 5.75 & 2.46 & 1.56 \\
 $B_c^- \to \overline B_s^0 \rho^-$    & 2.3  & 7.2  & 4.41  & 1.38 & 3.86 \\
 $B_c^- \to \overline B_s^{*0} \pi^-$  & 2.1  & 6.5  & 5.08 & 1.58 & 1.23 \\
 $B_c^- \to \overline B_s^{\ast\,0} \rho^-$ & 11  & 20.2 & 14.8  & 10.8 & 16.8 \\
 $B_c^- \to \overline B_s^0 K^-$       & 0.29  & 1.06 & 0.41  & 0.21 & 0.17  \\
 $B_c^- \to \overline B_s^{\ast\,0} K^-$    & 0.13  & 0.37 & 0.29 & 0.11 &  0.13  \\
 $B_c^- \to \overline B_s^{\ast\,0} K^{\ast\,-}$ & 0.50  & -    & -    & -& 1.14 \\
\hline
 $B_c^- \to \overline B^0 \pi^-$       & 0.20  & 1.06  & 0.32  & 0.10 & 0.10 \\
 $B_c^- \to \overline B^0 \rho^-$      & 0.20  & 0.96  & 0.59  & 0.13 & 0.28 \\
 $B_c^- \to \overline B^{\ast\,0} \rho^-$   & 0.30    & 2.57  & 1.17  & 0.67 & 0.89 \\
\hline
 $B_c^- \to B^-  K^0$       & 0.38  & 1.98  & 0.66 & 0.23 & 0.27   \\
 $B_c^- \to B^- K^{\ast\,0}$    & 0.11  & 0.43  & 0.47   & 0.09 & 0.32 \\
 $B_c^- \to B^{\ast\,-} K^{\ast\,0}$ & 0.32  & 1.67  & 0.97 & 0.82 & 1.70 \\
\hline\hline
\end{tabular}
\end{table}

\begin{table}[t]
\caption{\label{tab:Bc-nonlep2}
         The same as of table \ref{tab:Bc-nonlep1}.}
\begin{center}
\begin{tabular}{|l|r|r|r|r|}
\hline
 Mode &  \cite{last} & \cite{Chang:2001pm} &\cite{Kiselev:2001zb} &
\cite{LopezCastro:2002ud}  \\
\hline
 $B_c^-\to h_c\, \pi^-$          & 0.11    & 0.05      & 1.60   & -       \\
 $B_c^-\to \chi_{c0}\, \rho^-$   & 0.13    & 0.072     & 3.29   &-        \\
 $B_c^-\to h_c\, \rho^-$         & 0.25    & 0.12      & 5.33   &-        \\
 $B_c^-\to \chi_{c2}\, \rho^-$   & 0.12    & 0.051     & 3.20   & 0.023   \\
\hline
\end{tabular}
\end{center}
\end{table}

From the tables we observe that our
results are generally close to the QCD sum rule results of \cite{KKL,exBc} and
the constituent quark model results of \cite{Chang:1992pt,Faust,AbdEl-Hady:1999xh}
for the $b \to c$ induced decays. In
exception are the $(b \to c ; c \to (s,d))$ results of \cite{Chang:1992pt}
which are considerably smaller than our results, and smaller than the
results of the other model calculations. Summing up the exclusive
contributions one obtains a branching fraction of $8.8\%$. Considering the
fact that the $b \to c$ contribution to the total rate is expected to be
about 20\% \cite{Brambilla:2004wf} this leaves plenty of room for
nonresonant multibody decays

For the $c \to s$ induced decays our branching ratios are considerably
smaller than those predicted by QCD sum rules \cite{KKL,exBc} but are
generally close to the other constituent quark model results. When we sum
up our exclusive branching fractions we obtain a total branching ratio of
27.6\% which has to be compared with the 70\% expected for the $c \to s$
contribution to the total rate \cite{Brambilla:2004wf}. The sum rule model
of \cite{KKL,exBc} gives a summed branching fraction of 73.4\% for the
$c \to s$ contribution, i.e. the model of \cite{KKL,exBc} predicts that the
exclusive channels pretty well saturate the $c \to s$ part of the total rate.

\section{Conclusions}

In the coming few years one can expect large data samples on
exclusive $B_c$ decays at the TEVATRON and at the LHC. We are looking forward to a
comparison of our model results with the upcoming experimental data.



\end{document}